\documentclass[
 reprint,
 superscriptaddress,
 amsmath,amssymb,
 aps,
prl,
]{revtex4-2}

\usepackage{graphicx}% Include figure files
\usepackage{hyperref}% add hypertext capabilities
\usepackage{siunitx}
\usepackage{xspace}
\usepackage{xcolor}
\usepackage{cancel}
\usepackage{tikz}
\usepackage{ulem}
\usepackage{mathtools}
\DeclarePairedDelimiter{\ceil}{\lceil}{\rceil}

\newcommand{\ket}[1]{\ensuremath{\lvert #1 \rangle}\xspace}%

\newcommand{\citesm}[1]{~\cite{sm}}

\makeatletter
\def\maketitle{
\@author@finish
\title@column\titleblock@produce
\suppressfloats[t]}
\makeatother

\begin{document}

\newcommand{\partitle}[1]{\subsection*{#1}}

\newcommand{\papertitle}{Observation of Hilbert-space fragmentation and fractonic excitations in two-dimensional Hubbard systems}

\title{\papertitle{}}

\author{Daniel Adler}
    \affiliation{Max-Planck-Institut f\"{u}r Quantenoptik, 85748 Garching, Germany}
    \affiliation{Munich Center for Quantum Science and Technology (MCQST), 80799 Munich, Germany}
\author{David Wei}
    \affiliation{Max-Planck-Institut f\"{u}r Quantenoptik, 85748 Garching, Germany}
    \affiliation{Munich Center for Quantum Science and Technology (MCQST), 80799 Munich, Germany}
\author{Melissa Will}
    \affiliation{Munich Center for Quantum Science and Technology (MCQST), 80799 Munich, Germany}
    \affiliation{Technische Universit\"{a}t M\"{u}nchen, TUM School of Natural Sciences, 85748 Garching, Germany}
\author{Kritsana Srakaew}
    \affiliation{Max-Planck-Institut f\"{u}r Quantenoptik, 85748 Garching, Germany}
    \affiliation{Munich Center for Quantum Science and Technology (MCQST), 80799 Munich, Germany}
\author{Suchita Agrawal}
    \affiliation{Max-Planck-Institut f\"{u}r Quantenoptik, 85748 Garching, Germany}
    \affiliation{Munich Center for Quantum Science and Technology (MCQST), 80799 Munich, Germany}
\author{Pascal~Weckesser}
    \affiliation{Max-Planck-Institut f\"{u}r Quantenoptik, 85748 Garching, Germany}
    \affiliation{Munich Center for Quantum Science and Technology (MCQST), 80799 Munich, Germany}
\author{Roderich Moessner}
    \affiliation{Max-Planck-Institut für Physik komplexer Systeme, 01187 Dresden, Germany}
\author{Frank Pollmann}
    \affiliation{Munich Center for Quantum Science and Technology (MCQST), 80799 Munich, Germany}
    \affiliation{Technische Universit\"{a}t M\"{u}nchen, TUM School of Natural Sciences, 85748 Garching, Germany}
\author{Immanuel Bloch}
    \affiliation{Max-Planck-Institut f\"{u}r Quantenoptik, 85748 Garching, Germany}
    \affiliation{Munich Center for Quantum Science and Technology (MCQST), 80799 Munich, Germany}
    \affiliation{Fakult\"{a}t f\"{u}r Physik, Ludwig-Maximilians-Universit\"{a}t, 80799 Munich, Germany}
\author{Johannes Zeiher}
	\affiliation{Max-Planck-Institut f\"{u}r Quantenoptik, 85748 Garching, Germany}
    \affiliation{Munich Center for Quantum Science and Technology (MCQST), 80799 Munich, Germany}

\date{\today}

\begin{abstract}
The relaxation behaviour of isolated quantum systems taken out of equilibrium is among the most intriguing questions in many-body physics.
%
%Usually, systems are expected to return to thermal equilibrium 
Quantum systems out of equilibrium typically relax to thermal equilibrium states by scrambling local information and building up entanglement entropy. However, kinetic constraints in the Hamiltonian can lead to a breakdown of this fundamental paradigm due to a fragmentation of the underlying Hilbert space into dynamically decoupled subsectors in which thermalisation can be strongly suppressed. Here, we experimentally observe Hilbert space fragmentation (HSF) in a two-dimensional tilted Bose-Hubbard model. Using quantum gas microscopy, we engineer a wide variety of initial states and find a rich set of manifestations of HSF involving bulk states, interfaces and defects, i.e.,  $d = 2$,\ 1 and 0 dimensional objects. Specifically, uniform initial states with equal particle number and energy differ strikingly in their relaxation dynamics.  Inserting controlled defects on top of a global, non-thermalising chequerboard state, we observe highly anisotropic, sub-dimensional dynamics, an immediate signature of their fractonic nature. An interface between localized and thermalising states in turn displays dynamics depending on its orientation. Our results mark the first observation of HSF beyond one dimension, as well as the concomitant direct observation of fractons, and pave the way for in-depth studies of microscopic transport phenomena in constrained systems.

\end{abstract}

\maketitle

%%%%%%%%%%%%%%%%%%%%%%%%%%%%%%%%%%%%%%%%%%%%
%               Introduction               %
%%%%%%%%%%%%%%%%%%%%%%%%%%%%%%%%%%%%%%%%%%%%
\begin{figure}[h!]
    \centering
    \includegraphics{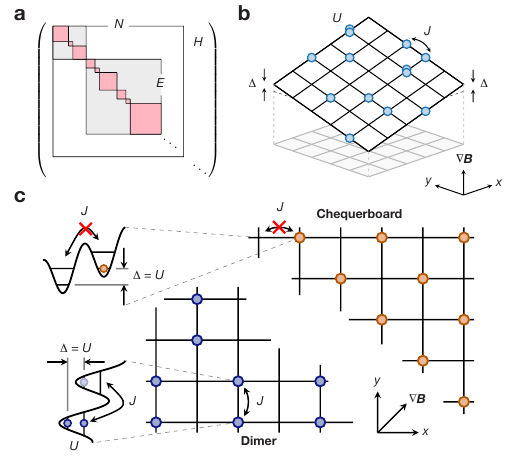}
    \caption{
    \textbf{Hilbert space fragmentation and schematic of the experiment.}
        \textbf{a} The Hilbert space consists of sectors with fixed energy and particle number (conserved quantities), $(E,N)$ (grey squares). Adding dynamical constraints to the system, these sectors fragment into decoupled Krylov subsectors (pink squares).
        \textbf{b} Our system is described by a tilted Bose-Hubbard model with a diagonal tilt along the $x+y$-direction tuned to resonance with the interactions, $\Delta = U$. Tilt and interaction energy are much larger than the tunnel coupling $J$.
        \textbf{c} Dimer (blue points, lower left) and chequerboard state (orange points, upper right) and first-order processes exemplifying the presence (absence) of density-assisted resonant couplings in the lattice for the dimer (chequerboard) state.
        }
    \label{fig:1}
\end{figure}
The eigenstate thermalisation hypothesis (ETH) expresses the  notion of thermalisation in closed quantum systems by stating that eigenstates produce expectation values for local observables that are consistent with those of a thermal ensemble, and thus lose all memory of the initial states during relaxation~\cite{Polkovnikov2011,DAlessio2016,Deutsch2018}.
Recently, several mechanisms have been delineated, where systems defy such thermalising dynamics and ETH, such as integrability in one-dimensional systems~\cite{Essler2016,Bertini2021}, many-body localisation in models with quenched disorder~\cite{Nandkishore2015, Altman2018,Abanin2019}, or the emergence of many-body scars for specific initial settings~\cite{Serbyn2021, Moudgalya2022a}.
Another mechanism is the emergence of kinetic constraints connected with Hilbert space fragmentation (HSF)~\cite{Garrahan2018,Lan2018,Khemani2020,Sala2020,Moudgalya2020,Khudorozhkov2022,Moudgalya2022,Moudgalya2022a}.
In systems exhibiting HSF, a hierarchy of conservation laws exists.
First, the full Hilbert space can be divided into (polynomially many) subspaces characterized by global quantum numbers such as particle number or dipole moment.
In the corresponding subspaces with constant quantum numbers, local kinetic constraints lead to a further fragmentation of the Hilbert space into exponentially many smaller subsectors, the so-called ``Krylov" subsectors, which cannot be characterised by simple quantum numbers.
All states in a single Krylov sector are, by definition, dynamically connected, i.e., can be reached through unitary time evolution with the Hamiltonian~\cite{Khemani2020,Sala2020}.
One striking consequence of HSF is the possible existence of fragments containing specific states that evade thermalisation because of the underlying kinetic constraints and the small size of the associated Krylov sector.

Another particularly interesting consequence of constrained dynamics is the potential emergence of fractons that exhibit restricted mobility~\cite{Nandkishore2019,Pretko2020,Will2023}. 
Fractons can either be immobile under local Hamiltonian dynamics, or exhibit subdimensional dynamics, such as propagation in an effectively one-dimensional subspace of two-dimensional space~\cite{Will2023}, as well as anomalous diffusion~\cite{Feldmeier2020, Morningstar2020, Gromov2020}.
Previous theoretical studies~\cite{Nandkishore2019,Pai2019,Pretko2020,Sala2020,Khemani2020,Moudgalya2022a} have also related the emergence of fractons to gauge theories associated with local conservation laws and to topological defects in elasticity theory~\cite{Pretko2018}.
Relaxation of systems exhibiting fractonic excitations is expected to be strongly impeded, leading to non-ergodic behaviour and strongly temperature dependent transport dynamics~\cite{Pretko2020,Morningstar2020}.
Subdiffusive transport in the tilted Fermi-Hubbard model has recently been observed experimentally~\cite{Guardado-sanchez2020}.
Related theoretical work has connected the emerging subdiffusive hydrodynamic behaviour with the presence of kinetic constraints ~\cite{Zhang2020,Gromov2020}.
The kinetic constraints underlying HSF have been experimentally probed directly in one-dimensional tilted Hubbard chains~\cite{Scherg2021, Kohlert2023}.
These systems exhibit dipole-moment conservation for strong interactions, as a consequence of the interplay between interaction and tilt energy~\cite{Sachdev2002, Simon2011, Pielawa2011, Khemani2020, Sala2020}.
Recently, state-specific relaxation behaviour in systems with HSF was also observed in Rydberg arrays of one-dimensional kinetically constrained in the facilitation regime~\cite{Zhao2024} and quantum ladders realised with a superconducting quantum processor~\cite{Wang2024}.
The evasion of thermalisation tends to depend, frequently qualitatively, on the underlying dimensionality, particularly famously for both integrable or disorder-localised systems.
Consequently, the question then naturally arises what are the hallmarks of non-ergodicity in {\it higher-dimensional} HSF. 
%

%%%%%%%%%%%%%%%%%%%%%%%%%%%%%%%%%%%%%%%%%%%%
%               Here, we show               %
%%%%%%%%%%%%%%%%%%%%%%%%%%%%%%%%%%%%%%%%%%%%
Here we investigate this question for a two-dimensional tilted Bose-Hubbard model, where we study the non-equilibrium dynamics due to HSF in bulk ($d=2$), interface ($d=1$) and point-like defect ($d=0$) dynamics, and find a rich and interrelated phenomenology. 
Our experiments leverage the single-site control achievable in a quantum gas microscope to prepare specific initial product states in different Krylov sectors and measure their dynamics after a quantum quench.
We find dramatically different relaxation dynamics of a chequerboard state and a dimer state (see Fig.~\ref{fig:1}c), which are characterised by the same quantum numbers but are part of different Krylov subsectors.
Moreover, we prepare and dynamically track defects on top of the otherwise immobile chequerboard state.
Our measurements reveal the fractonic nature of such defects, which manifests itself as strongly constrained, sub-dimensional motion along a one-dimensional manifold in the two-dimensional system.
Finally, we prepare an interface between a chequerboard state and a dimer state and observe strongly asymmetric dynamics across the interface consistent with the fractonic nature of the excitations.
%

%%%%%%%%%%%%%%%%%%%%%%%%%%%%%%%%%%%%%%%%%%%%
%     Physical background (Fig. 1)         %
%%%%%%%%%%%%%%%%%%%%%%%%%%%%%%%%%%%%%%%%%%%%
The tilted Bose-Hubbard model has been studied in a number of works theoretically~\cite{Sachdev2002,Pielawa2011,Lake2022} and experimentally, focusing on the interesting ground-state phases~\cite{Simon2011,Kim2024} or emerging long-range tunneling dynamics~\cite{Meinert2014}.
The corresponding Hamiltonian is given by
\begin{align}
    \hat{H} = &- J \sum_{\left\langle i, j \right\rangle} \hat{a}_i^\dagger \hat{a}_j + \frac{U}{2} \sum_{i,j} \hat{n}_{i,j} \left( \hat{n}_{i,j} - 1 \right) \notag \\ &+ \Delta \sum_{i,j} \left(i + j \right) \hat{n}_{i,j},
    % + \sum_{i,j} V_{i,j} \hat{n}_{i,j},    
\end{align}
where the sum over $\left\langle i, j \right\rangle$ runs over all nearest-neighbor sites.
The tunnel coupling between two sites in the lattice is denoted as $J$, and the interaction energy of two bosons occupying the same site is $U$.
Applying a strong tilt along the diagonal of the lattice with $\Delta\gg J$ introduces dynamical constraints.
A particularly interesting regime is reached in the limit $U/J \gg 1$ with resonant tilt $\Delta = U$.
Here, both particle number $N$ and the sum of tilt and interaction energy, $E = \sum_i \Delta_i + U_i$, with the sum running over all sites $i$, are approximately conserved globally such that sectors with fixed quantum numbers $(E,N)$ emerge~\cite{Will2023}, see Fig.~\ref{fig:1}a. 
Additionally, atoms can only couple resonantly to already occupied sites and are thus subject to strong dynamical constraints.
Retaining only terms up to and including second order in $J/U$, these constraints were recently shown to result in HSF~\cite{Will2023}.
In particular, HSF can be observed in the first order in this model, which experimentally allows to access longer timescales compared to other models exhibiting HSF based on second-order processes~\cite{Khemani2020,Sala2020}.
Two states of a single sector with fixed $(E,N)$ expected to exhibit dramatically different thermalisation behaviour are the chequerboard state and the dimer state shown in Fig.~\ref{fig:1}c.
In the chequerboard state, isolated atoms are not coupled to neighbouring sites, and they are expected to remain frozen and retain memory of the initial density pattern.
This contrasts with the dimer state, which is characterised by neighbouring pairs and thus features resonances $\Delta=U$ that can facilitate dynamics and thus lead to a relaxation of the initial density pattern.

%%%%%%%%%%%%%%%%%%%%%%%%%%%%%%%%%%%%%%%%%%%%
%             Experimental system          %
%%%%%%%%%%%%%%%%%%%%%%%%%%%%%%%%%%%%%%%%%%%%
\begin{figure}[t!]
    \centering
    \includegraphics{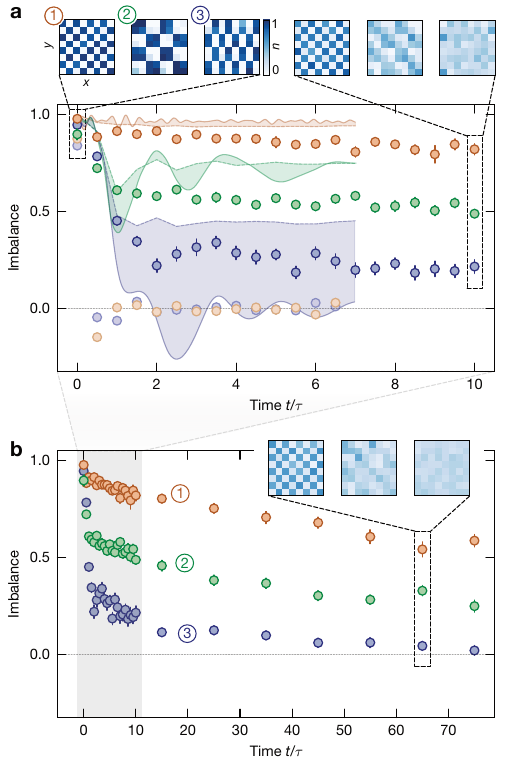}
    \caption{
    \textbf{Relaxation of the imbalance for different initial states.}
        \textbf{a} Short-time evolution of the imbalance. Imbalance of different initial states as a function of evolution time in units of the hopping timescale $\tau$ for the chequerboard (orange), squares (green), and dimer (blue) initial states with (circles) and without (desaturated circles) applied tilt. The imbalance in the case without tilt quickly decays to zero regardless of the initial state, whereas the decay strongly depends on the initial state once the tilt is applied, a clear signature of HSF. Insets: Average densities corresponding to the respective states at the indicated times in the $8 \times 8$ sites region of interest (ROI).
        The shaded, coloured areas denote the areas between theoretical calculations under imperfect (dashed lines) and perfect (solid lines) conditions \citesm{}.
        \textbf{b} Imbalance for longer evolution times. The grey shaded area highlights the data points shown in Fig.~\ref{fig:2}a. All error bars denote the standard error of the mean (s.e.m.).
        }
    \label{fig:2}
\end{figure}
We start our experiments by preparing a near-unity-filled Mott insulator of about $200$ bosonic $^{87}\mathrm{Rb}$ atoms in the $\ket{F = 1, m_F = -1}$ ground state in a single slice of a vertical optical lattice.
In the two-dimensional plane, we set the Hubbard parameters by controlling the depth of a two-dimensional folded horizontal lattice~\cite{wei2023}.
We use a digital micromirror device (DMD) to reduce the harmonic confinement induced by the optical lattice beams and realise approximately homogeneous trapping conditions~\cite{wei2022}.
We then exploit single-site addressing~\cite{weitenberg2011, fukuhara2013} to prepare different initial states in sectors with fixed energy and particle number $(E,N)$.
Next, we adiabatically ramp up a potential gradient using a magnetic field to the resonance condition $\Delta \approx U$ and then quench the lattice depth to $U/J \gg 1$ \citesm{}, initiating dynamics.
After a variable evolution time, we rapidly ramp up the lattice to freeze the dynamics and then record a fluorescence image of the parity-projected occupation per lattice site~\cite{sherson2010}.
%

%%%%%%%%%%%%%%%%%%%%%%%%%%%%%%%%%%%%%%%%%%%%
%             Fig. 2: Relaxation           %
%%%%%%%%%%%%%%%%%%%%%%%%%%%%%%%%%%%%%%%%%%%%
In a first set of measurements, we aim to directly show the emergence of HSF via the vastly different dynamics of different initial states in our model~\cite{Will2023}.
We prepare the chequerboard state, the dimer state, and also the ``squares" state, a chequerboard-like arrangement where four atoms and four empty sites, respectively, form the building blocks of a larger chequerboard-like structure.
For perfect initial state preparation, all of these states have the same energy and particle number.
The chequerboard state is part of a small fragment, dynamically disconnected from all other states, and thus frozen, whereas the dimer state is part of the largest fragment of the Hilbert space.
The squares state is expected to lie in-between, i.e., is part of a larger but not the largest fragment.
To probe the relaxation behavior for each pattern, we evaluate the imbalance defined as
\begin{equation}
    \mathcal{I} = \frac{N_o - N_u}{N_o + N_u}
\end{equation}
where $N_o, N_u$ are the parity-projected, detected number of atoms on initially occupied and unoccupied sites respectively.
The imbalance captures the degree to which the system retains a memory of the initially prepared pattern.
\begin{figure}[ht!]
    \centering
    \includegraphics{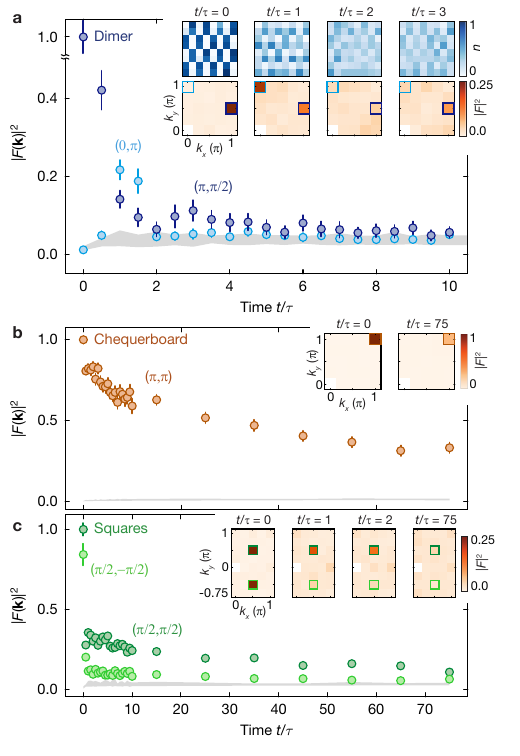}
    \caption{
        \textbf{Microscopic study of relaxation.}
        \textbf{a} Fourier analysis of the average densities for the dimer state. The $(\pi, \pi/2)$ Fourier component corresponding to the dimer state (dark blue) shows a fast decay, whereas the $(0, \pi)$ Fourier component for the CDW along the vertical direction (light blue) increases before decreasing again, corresponding to the first hopping processes.
        \textbf{b} Fourier analysis for the chequerboard state. The $(\pi, \pi)$ component decays only slightly and remains the dominant component.
        \textbf{c} Fourier analysis of the squares state. Both $(\pi/2,- \pi/2)$ the and the $(\pi/2, \pi/2)$ components decay quickly. The $(\pi/2,- \pi/2)$ component, which describes decay in the direction of the equipotential lines, decays to a lower value and quickly becomes indistinguishable from the background, while the $(\pi/2, \pi/2)$ component is above the background even at late times.
        For all initial states, all other components fall in between the grey-shaded areas describing the homogeneous background.
        Insets show the discrete 2D Fourier transforms (orange colormap) of the average densities (blue colormap) for selected times. The coloured rectangles highlight the Fourier components shown in the plots.
        Error bars denote the s.e.m.
        }
    \label{fig:3}
\end{figure}
\begin{figure*}
    \centering
    \includegraphics{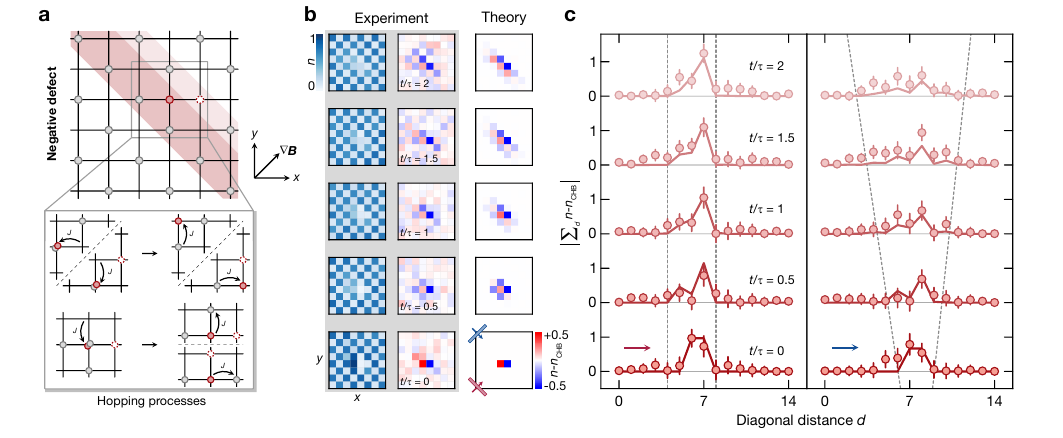}
    \caption{
        \textbf{Dynamics of fractonic excitations (negative defect).}
        \textbf{a} Schematic of the spreading of the defect atom (filled red circle) to first order, which can move by forming doublons (dark red shaded area). The defect hole (dashed circle) can also move, facilitated by the defect atom (light red shaded area).
        \textbf{b} Anisotropic spreading of the defect. The spreading of the defect can be observed in the occupations (due to parity projection) and more clearly in the difference plots when subtracting the time-evolved reference chequerboard background. Theory calculations are shown for comparison \citesm{}. The motion of the defect atom is restricted to a narrow stripe, along which it can move in either direction.
        \textbf{c} Diagonal sums over the reference-subtracted occupations. Orthogonal to the equipotential lines (left), the defect atom cannot spread, whereas parallel to the equipotential lines (right), it spreads in both directions. Dashed grey lines are guides to the eye. Solid lines are theory calculations, rescaled by the initial experimental defect population \citesm{}. The coloured arrows show the direction of the summation of the diagonals, as indicated by the insets in the bottom, rightmost plot in Fig.~\ref{fig:4}b.
        Error bars denote the s.e.m.
    }
    \label{fig:4}
\end{figure*}
\begin{figure*}
    \centering
    \includegraphics{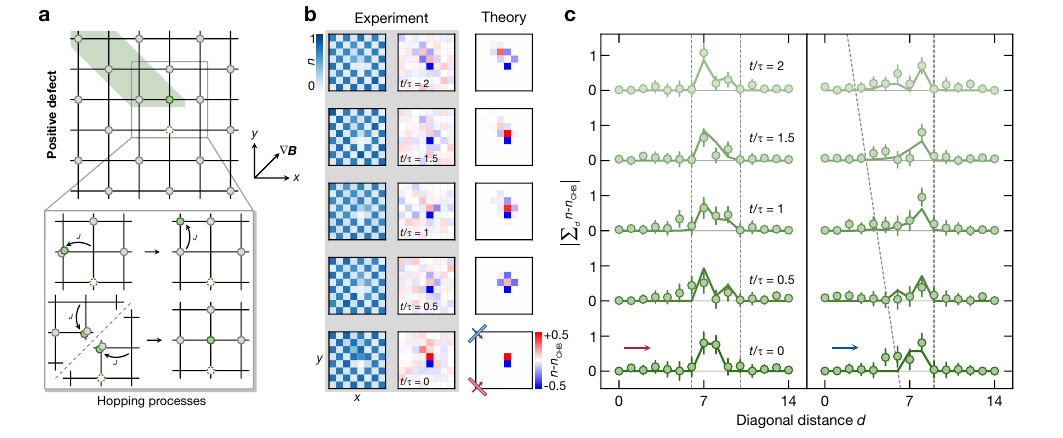}
    \caption{
        \textbf{Dynamics of fractonic excitations (positive defect).}
        \textbf{a}  Schematic of the spreading of the defect atom (filled green circle) to first order, which can move by forming doublons (dark green shaded area). On short timescales (to first order), the associated defect hole (dashed circle) is stuck.
        \textbf{b}  Anisotropic spreading of the defect analyzed in analogy to Fig.~\ref{fig:4}.
        \textbf{c} Sums over the reference-subtracted occupations, analyzed and compared to theory in analogy to Fig~\ref{fig:4}. Compared with the negative defect, the positive defect shows a clearly asymmetric propagation on the sub-dimensional manifold.
        }
    \label{fig:5}
\end{figure*}
\begin{figure}[ht!]
    \centering
    \includegraphics{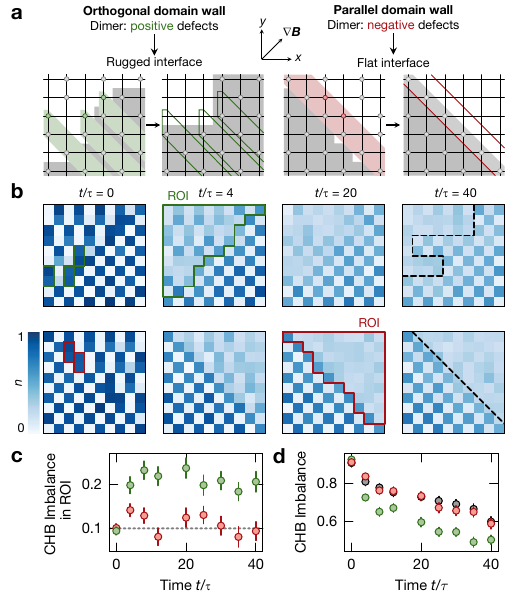}
    \caption{
        \textbf{Domain wall dynamics between chequerboard and dimer state.}
        \textbf{a} Depending on the interface orientation relative to the tilt (orthogonal, left; parallel, right), the dimers act as positive (green circles) or negative (red circles) defects in the chequerboard (grey area). At the interface, the upper atom in the dimers can propagate, whereas the lower atom stays at its original site.
        \textbf{b} Mean density as a function of evolution time for the central $10 \times 10$ sites. We observe the emergence of a larger chequerboard area for the orthogonal orientation (upper row), indicated by the black dashed line at $t/\tau = 40$. For the parallel orientation, the chequerboard remains intact.
        \textbf{c} Evaluating the data in the green and red triangular ROIs illustrated in Fig.~\ref{fig:6}b, we find the overlap with the chequerboard to remain constant for parallel orientation (red) while we observe an increase for the orthogonal orientation (green). 
        \textbf{d} For orthogonal interface orientation (green), the imbalance in the chequerboard half of the system decays. For parallel orientation (red), the imbalance stays consistent with the reference measurement of a pure chequerboard (grey).
        All error bars denote the s.e.m.
        }
    \label{fig:6}
\end{figure}

Tracking the evolution of the states in Fig.~\ref{fig:2}, we find that for the chequerboard, the imbalance is finite and large even for the longest evolution times up to $t/\tau = 80$, where $\tau = \hbar/J$ denotes the time scale associated with tunnelling in our experiment.
We attribute the initial small decay of the chequerboard imbalance within a few $\tau$ to imperfect preparation of the initial state and higher-order processes~\cite{Will2023}.
In contrast, the dimer state initially decays much faster to a strikingly lower imbalance, which then only slowly decays towards zero for the longest evolution times.
The imbalance of the squares pattern is found to lie approximately between the two extremal cases.
Interestingly, analysing the density at the largest evolution times, we observe that the residual imbalance for the squares pattern is due to a larger-scale structure in the density formed in particular by sites that are inaccessible for the atoms due to the presence of kinetic constraints. 
%

%%%%%%%%%%%%%%%%%%%%%%%%%%%%%%%%%%%%%%%%%%%%
%       Fig. 3: Microscopic picture        %
%%%%%%%%%%%%%%%%%%%%%%%%%%%%%%%%%%%%%%%%%%%%
The relaxation of the dimer state occurs microscopically through a resonant three-site subsystem that is initially connected via first-order tunnelling, which effectively allows the dimers to flip their orientation.
These processes are clearly visible in the time evolution of the density; see Fig.~\ref{fig:3}a, where the initial dimer pattern evolves into a stripe-like pattern resembling a charge density wave (CDW) before evolving back into the dimer pattern.
This characteristic relaxation behaviour also becomes apparent in a Fourier analysis of the density.
Following a quick initial decay of the initial Fourier component $(\pi,\pi/2)$ characterising the dimer state, we observe the growth of the $(0, \pi)$ component corresponding to a charge density wave (CDW) along the $y$-direction.
Subsequently, we also observe a small revival of the initial dimer pattern during the relaxation dynamics, both in density and Fourier component; see Fig.~\ref{fig:3}a.
In stark contrast, for the chequerboard state, the $(\pi,\pi)$ component remains the dominant Fourier component at all times and exhibits a fast initial decay followed by a slow decrease at long times.
For the squares initial state, two Fourier components with orthogonal orientation are relevant.
First, a fast relaxation occurs within the squares, whereas the coupling between different squares leads to a further slow decay of the $(\pi/2, \pi/2)$ and $(\pi/2, -\pi/2)$ components, as shown in Fig.~\ref{fig:3}c, consistent with the slow relaxation of the imbalance.
Here, the $(\pi/2, -\pi/2)$ component exhibits a faster decay compared to the $(\pi/2, \pi/2)$ component and becomes consistent with the ``background" of all other components at late times.
In contrast, the $(\pi/2, \pi/2)$ component is above the background level at all times.
This is due to the faster decay of the initial state along the direction of the equipotential lines, which corresponds to the $(\pi/2, -\pi/2)$ component.
Orthogonal to this direction, as described by the $(\pi/2, \pi/2)$ component, the kinetic constraints inhibit this decay, as is also visible in the inset in Fig.~\ref{fig:2}b.
%

%%%%%%%%%%%%%%%%%%%%%%%%%%%%%%%%%%%%%%%%%%%%
%             Fig. 4 Fractons              %
%%%%%%%%%%%%%%%%%%%%%%%%%%%%%%%%%%%%%%%%%%%%
After establishing the strong dependence of the observed dynamics on the initially prepared state, we aim to study the dynamics of excitations on top of the fragmented states.
Due to the kinetically constrained dynamics, defects prepared on top of the chequerboard state are expected to exhibit fractonic behavior~\cite{Will2023}.
To prepare ``negative" (``positive") defects, we displace one atom in the chequerboard state by one site such that its energy with respect to the tilt is decreased (increased).
As shown in Fig.~\ref{fig:4}a for the negative defect, the displacement leads to the emergence of new resonant processes, rendering the defect mobile with a subsequent dynamical evolution.
We make two striking observations when tracking the dynamics of negative and positive defects following a quench to finite tunnelling.
First, both types of defects are confined to movements in a one-dimensional subspace along the equipotential lines on the lattice grid, see Fig.~\ref{fig:4}c and Fig.~\ref{fig:5}c, left panel, respectively.
Second, within this one-dimensional subspace, the positive defect propagates asymmetrically to only one side, which can be understood from the presence of a blocked site in the direct vicinity of the prepared defect, see Fig.~\ref{fig:5}c, right panel.
For the positive defect, the hole associated with the displaced atom is immobile to first order and, for the times studied here, remains at the site where it was originally created.
We observe the asymmetric propagation also directly in the average occupation as a reduction of the chequerboard contrast in the direction where the defect can move, see Fig.~\ref{fig:4}b and Fig.~\ref{fig:5}b.
The sub-dimensional propagation for both positive and negative defects and the unidirectional motion of the positive defect are strong indications that the prepared defects indeed exhibit the expected fractonic properties.
%

%%%%%%%%%%%%%%%%%%%%%%%%%%%%%%%%%%%%%%%%%%%%
%             Fig. 5: Interface            %
%%%%%%%%%%%%%%%%%%%%%%%%%%%%%%%%%%%%%%%%%%%%
In a final set of measurements, we study the relaxation of an interface between the mobile dimer state and the immobile chequerboard state.
Such a measurement directly probes the impact of kinetic constraints for the underlying defects on transport characteristics in systems exhibiting HSF. 
Interestingly, we observe strikingly different dynamics depending on the alignment of the interface along or orthogonal to the equipotential surfaces.
If the interface is aligned with the equipotential surface, the chequerboard state remains stable, whereas the dimer rapidly decays.
Importantly, the interface stays intact, which indicates that -- consistent with the sub-dimensional character of the defects -- no transport occurs across the boundary.
Conversely, if the interface is oriented orthogonal to the equipotential lines, a chequerboard structure starts building up even on the sites initially prepared with a dimer pattern.
These observations can be directly explained by the type and location of defects initially injected into the system, in combination with the strongly asymmetric fractonic character of single defects, see Fig.~\ref{fig:6}a.
In the vicinity of the interface, the mobile atom in each initially prepared dimer will propagate into the chequerboard region.
The remaining atoms in these dimers form a chequerboard pattern, effectively increasing the overall area of the then immobile chequerboard.
This is evidenced by the increase of the chequerboard imbalance evaluated in the half of the system initially prepared in the dimer state, see Fig.~\ref{fig:6}c.
Simultaneously, as shown in Fig.~\ref{fig:6}d, the chequerboard imbalance within the other half of the system drops significantly lower than for an independent reference chequerboard measurement without interface or for the interface orientation parallel to the equipotential lines, due to the influence of the mobile defects.
%

%%%%%%%%%%%%%%%%%%%%%%%%%%%%%%%%%%%%%%%%%%%%
%            Conclusion and Outlook        %
%%%%%%%%%%%%%%%%%%%%%%%%%%%%%%%%%%%%%%%%%%%%
In summary, we have demonstrated the emergence of HSF in a two-dimensional tilted Bose-Hubbard model as a consequence of strong kinetic constraints.
Our measurements mark the first comprehensive study of HSF in two-dimensional systems through their thermalisation properties, which differ strikingly for different initial states.
Our results immediately spark follow-up questions, such as whether the tilted Bose-Hubbard model is weakly or strongly fragmented.
Such a study would require measurements for much larger system sizes and a finite-size scaling analysis of the final imbalance, or the preparation of a number of other initial states from one ($E,N$) subsector.
It would also be interesting to study the predicted diffusive and subdiffusive behaviour of negative and positive defects~\cite{Will2023}, which would be possible in larger systems with longer coherence times.
Furthermore, the detailed understanding of the complex dynamics emerging at the interface between states from different fragments remains open and is left for further work, stressing that numerical simulations are limited to evolution times much smaller than those accessible in our experiment.
In future work, it may also be interesting to study HSF in open systems experimentally by adding controlled dissipation~\cite{Li2023}, or explore quantum HSF, where the Hilbert space fragments are characterised by entangled substates~\cite{Brighi2023}.

\begin{acknowledgments}
The authors thank Monika Aidelsburger, Sarang Gopalakrishnan, and David Huse for stimulating discussions.  
We acknowledge funding by the Max Planck Society (MPG) the Deutsche Forschungsgemeinschaft (DFG, German Research Foundation) under Germany's Excellence Strategy--EXC-2111--390814868 and through the DFG Research Unit FOR 5522 (project-id 499180199), as well as by the cluster of excellence
ct.qmat (EXC-2147--390858490).
This publication has also received funding under the Horizon Europe program HORIZON-CL4-2022-QUANTUM-02-SGA via the project 101113690 (PASQuanS2.1).
K. S. and S. A. acknowledge funding from the International Max Planck Research School (IMPRS) for Quantum Science and Technology.
F. P. acknowledges support from European Union’s Horizon 2020 research and innovation program under grant agreement No. 771537, the DFG TRR 360 - 492547816, and the Munich Quantum Valley, which is supported by the Bavarian state government with funds from the Hightech Agenda Bayern Plus.
P.W. acknowledges funding through the Walter Benjamin programme (DFG project 516136618).
\end{acknowledgments}

%%%%%%%%%%%%%%%%%%%%%%%%%%%%%%%%%%%%%%%%%
% References                            %
%%%%%%%%%%%%%%%%%%%%%%%%%%%%%%%%%%%%%%%%%

\bibliography{HSF}
\clearpage

%%%%%%%%%%%%%%%%%%%%%%%%%%%%%%%%%%%%%%%%%%%%%%%%%%%%%%%%%%%%%%%%%%%%%%%%%%%%%%%%
%                          SUPPLEMENTARY INFORMATION                           %
%%%%%%%%%%%%%%%%%%%%%%%%%%%%%%%%%%%%%%%%%%%%%%%%%%%%%%%%%%%%%%%%%%%%%%%%%%%%%%%%

\setcounter{equation}{0}
\setcounter{figure}{0}
\setcounter{table}{0}
\setcounter{section}{0}
\setcounter{subsection}{0}
\renewcommand{\theequation}{S\arabic{equation}}
\renewcommand{\thefigure}{S\arabic{figure}}
\renewcommand{\thetable}{S\arabic{table}}

\title{
    Supplementary Material for: \\
    \papertitle{}
}
\maketitle

\section{Experimental calibrations and data evaluation}

\subsection{Experimental details}

Here, we briefly describe the initial state preparation common to all measurements.
Experiments are performed in a single plane of a vertical 1D optical lattice. For the in-plane lattice, we use the folded lattice described in~\cite{wei2023}. Since the in-plane lattice is subject to disorder and harmonic confinement, we use a digital micromirror device (DMD) to shape the horizontal on-site potential, allowing us to achieve approximately homogeneous trapping depths and tunnelling energies throughout the system.
Employing a second DMD, we additionally project a tapered, rectangular box in the centre of this corrected system, to achieve reliable loading and high filling in a central area of about $15 \times 15$ lattice sites. \hfill \break
Starting from these Mott insulators (MI), in order to prepare the initial states of interest, we then perform local addressing over the entire area \cite{weitenberg2011, fukuhara2013}, while data analysis is performed in a smaller region of interest (ROI) of either $8 \times 8$ or $10 \times 10$ lattice sites at the centre of the system. Additionally, working with larger systems than the size of the ROI minimizes the influence of finite-size and boundary effects. 
With this preparation sequence, we achieve a filling of $0.88(2)$ per site on the addressed sites and a filling of $0.04(2)$ on the non-addressed sites in the ROI. These values are averaged over all datasets and initial configurations.

\subsection{Magnetic field gradient calibration}

\begin{figure}[t!]
    \centering
    \includegraphics{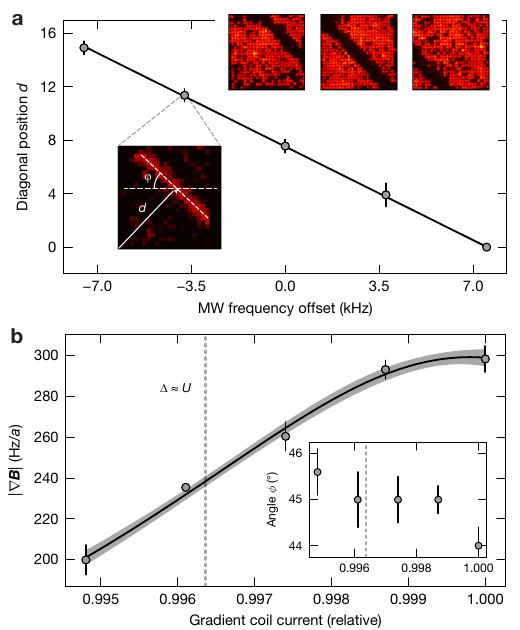}
    \caption{
        \textbf{Magnetic field gradient calibration.} 
        \textbf{a} Fitted center positions of the spatial profiles; gradient calibration (strength and orientation).
        Insets show exemplary single shots of the MW resonances for different centre frequencies. These shots are averaged and fitted with a 2D Gaussian to extract the gradient strength and orientation.
        \textbf{b} Gradient strength from the analysis in Fig.~\ref{fig:si:gradient-calibration}a at different gradient coil settings. The magnetic field is interpolated using Eq.~\ref{eq:mag_grad}.
        The inset shows that the angle does not change significantly even when tuning the gradient.
        The vertical, dashed lines indicate the operating point chosen based on the measurement in Fig.~\ref{fig:si:step-gradient}.
    }
    \label{fig:si:gradient-calibration}
\end{figure}

\begin{figure}
    \centering
    \includegraphics{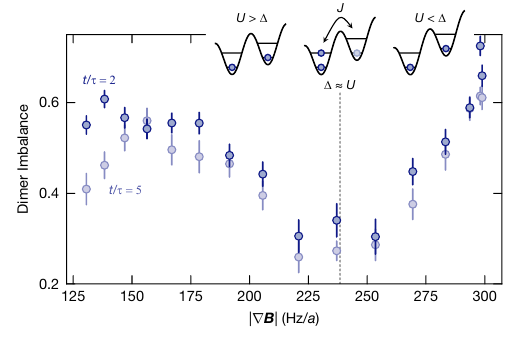}
    \caption{
        \textbf{Dimer imbalance for different gradient strengths.}
        Imbalance of the dimer at $t / \tau = 2$ (dark blue circles) and $t / \tau = 5$ (light blue circles) for different gradient strengths.
        The vertical line indicates the operating point chosen for all other measurements.
        Insets illustrate our expectations about the functional shape of the imbalance (see text) based on the possible processes.
        Error bars denote the standard error of the mean (s.e.m.).
    }
    \label{fig:si:step-gradient}
\end{figure}

The potential tilt in our experiments is realized by global magnetic fields, which allow us to induce the most homogeneous gradients.
We calibrate the magnitude and the orientation of the magnetic gradient using spatially resolved microwave (MW) spectroscopy on the magnetic-field sensitive transition between the $\ket{F = 1, m_F = -1}$ and $\ket{F = 2, m_F = -2}$ hyperfine ground states. \hfill \break
To this end, we prepare a large MI with all atoms in the $\ket{F = 1, m_F = -1}$ state.
We then adiabatically ramp up the magnetic field to its target configuration and perform narrow MW sweeps at variable centre frequencies. As a consequence, atoms are addressed resonantly within a narrow stripe subjected to the same magnetic field strength and flipped into the $\ket{F = 2, m_F = -2}$ state, which are then removed before imaging.
We fit a 2D Gaussian to these stripes of missing atoms, which allows us to map the field strength and gradient orientation versus their position, see Fig.~\ref{fig:si:gradient-calibration}a. \hfill \break
To be able to continuously vary the applied gradient strength, we use a combination of coils:
a single vertical gradient coil and a pair of vertical offset coils in Helmholtz configuration with reversed field polarity realize a quadrupole field near the plane of the atoms. For the initial calibration, we work with a fixed gradient coil setpoint and tune the vertical offset and additional in-plane offset fields such that the magnetic zero-point is at the location of the atoms; subsequently we shift the zero-point by a fixed amount using the in-plane offset fields, resulting in an in-plane gradient at the correct angle.
We then proceed to calibrating the gradient strength for various gradient coil setpoints as described above; for technical reasons we tune the gradient coil instead of the offset coils.
We interpolate between the calibrated values by fitting them with the function

\begin{equation}
    g \left( \Delta B \right) = \frac{g_0^2 r}{\sqrt{g_0^2 r^2 + \left( \Delta B + B_0 \right)^2}},
    \label{eq:mag_grad}
\end{equation}
where $r$ is the displacement of the magnetic field zero to the atoms, $g_0$ is the maximal gradient strength, $B_0$ describes background fields and $\Delta B$ is the change of the setpoint of the gradient coil. Since we only change $\Delta B$ by a few percent, we can assume $g_0 \left( B \right) = \mathrm{const.}$, which is also supported by the fact that the fit function describes the data well, as shown in Fig.~\ref{fig:si:gradient-calibration}b. \hfill \break
Based on this curve, we can then rescale the x-axis in Fig.~\ref{fig:si:step-gradient} and obtain an absolute value for the gradient strength.

\subsection{Hubbard parameters}

To extract the Hubbard parameters of our folded optical lattice~\cite{wei2023}, we make use of two methods.
First, we perform amplitude modulation (AM) spectroscopy to calibrate the lattice depth.
The results are then compared to a bandstructure calculation to obtain the values for the on-site interaction $U$ and the tunnelling energy $J$.
Here, we find $U / J = 21(2)$ with $U = h\times\SI{275(5)}{\hertz}$ and $J = h\times\SI{13(1)}{\hertz}$. The error bars arise from the uncertainty of the lattice depth calibration itself as well as the slightly anisotropic hopping along the two lattice axes~\cite{wei2023}. %\hfill \break
We can independently calibrate the Hubbard parameters using the quench dynamics of isolated dimers, see Fig.~\ref{fig:si:iso-dim} and below.
As a result, we extract $\tau = \hbar/J = \SI{10.0(3)}{\milli\second}$, equivalent to $J = h \times \SI{16.0(5)}{\hertz}$. Comparing again to our bandstructure calculation, this corresponds to $U / J = 17(1)$ with $U = h\times\SI{260(5)}{\hertz}$.
We attribute the deviations between these two calibrations to day-to-day drifts of the lattice beam alignment over the entire data taking period. \hfill \break
For data evaluation, we use $\tau = \SI{11}{\milli\second}$ for all datasets; motivated by the long data taking period of several days for a given data set.
Theory calculations (see below) are performed for $U /J = 18$ which is chosen as an intermediate value between the two calibrations.

\subsection{Tuning the gradient to resonance}

For the presented studies, it is important that the applied gradient matches the on-site interaction, that is, $\Delta = U$.
We benchmark the resonance location by measuring the dimer imbalance as a function of the gradient strength for various tunnelling times, as illustrated in Fig.~\ref{fig:si:step-gradient}. Here, we expect the strongest decay of the dimer imbalance, as defined in the main text, when the resonance condition is fulfilled.
For smaller gradients, we expect a slower drop in imbalance, whereas for much stronger gradients, we expect all processes to be off-resonant and no dynamics to occur at all, leading to high imbalance even at later times. \hfill \break
Our experimental results match the described expectation qualitatively.
To confirm that we are not accidentally probing at a time where the imbalance shows any $\Delta$-dependent oscillations, we probe for multiple fixed evolution times (up to $t/\tau = 40$), observing consistent behaviour for all of the chosen evolution times.
The resonance width is inherently limited by the finite tunnelling bandwidth and residual potential disorder.
Our chosen operation point is located at the centre of the resonance and exhibits the strongest decay, as marked by the vertical dashed line in Fig.~\ref{fig:si:step-gradient}. 
Based on our gradient calibration presented above and in Fig.~\ref{fig:si:gradient-calibration}b this point corresponds to a value of $\Delta = h\times\SI{238(3)}{\hertz}$. \hfill \break
Comparing to our independent bandstructure calculation, we find a qualitative agreement within $\SI{15}{\percent}$ to the value of $U$  for both calibration methods of the Hubbard parameters described above.
In particular, $U$ changes only very slowly with the lattice depth and varies by less than $J$ for our calibrations. As such, this gradient setpoint remains valid throughout all measurements.

\subsection{Data analysis}

All data, unless specified differently, is analyzed as explained in the following: we calculate the quantity of interest (imbalance, Fourier components, diagonal sums) on the individual experimental shots, then average over these results to obtain the data shown in the figures.
For the reference-subtracted defect occupations (middle columns in Fig.~\ref{fig:4}b, Fig.~\ref{fig:5}b), we subtract the densities averaged over all shots.
In order to calculate the imbalance in Fig.~\ref{fig:6}c and Fig.~\ref{fig:6}d, we choose the boundary of the respective ROI such that atoms that could be part of either the chequerboard or the dimer are counted as belonging to the dimer; this then yields the same number of atoms for both halves of the system and explains the unintuitive shape of the ROI shown in Fig.~\ref{fig:6}b.

\subsection{Fourier analysis}

To analyse the Fourier components of the average densities, we calculate the discrete Fourier transform according to

\begin{multline}
    F \left( \mathbf{k} \right) = \sum_{n = 0}^{N -1} \sum_{m = 0}^{M -1} a_{n,m} \exp \left( - 2 \pi i \left( \frac{n j}{N} + \frac{m l}{M} \right) \right) \\
    = \sum_{n = 0}^{N -1} \sum_{m = 0}^{M -1} a_{n,m} \exp \left( - i k_x n - i k_y m \right)
\end{multline}
with $a_{n,m}$ the average densities at site $n,m$ for a ROI of size $N \times M$ and $k_x = \frac{2 \pi j}{N}, k_y = \frac{2 \pi l}{M}$. Here, the index $j$ runs from $- \left( \frac{N - 1}{2} \right), ... \hspace{2pt}, 0, ... \hspace{2pt}, \frac{N - 1}{2}$ for odd $N$ and $\ceil*{\frac{N - 1}{2}}, ... \hspace{2pt}, 0 , ... \hspace{2pt}, \frac{N}{2}$ for even $N$ and analogously for $l$.
The value at $\left( k_x, k_y \right) = (0,0)$ is just the sum of the signal; it contains no additional relevant information and is thus neglected (white rectangles in the insets of Fig.~\ref{fig:3}). \hfill \break
Note that the discrete Fourier transform obeys point reflection symmetry, i.e., $F(\mathbf{k}) = F(\mathbf{-k})$.
In the main text, we therefore only plot the parts of the momentum space $\left( k_x, k_y \right)$ that contain non-redundant information.

\section{Extended data sets}

\subsection{Isolated dimer dynamics}

\begin{figure}
    \centering
    \includegraphics{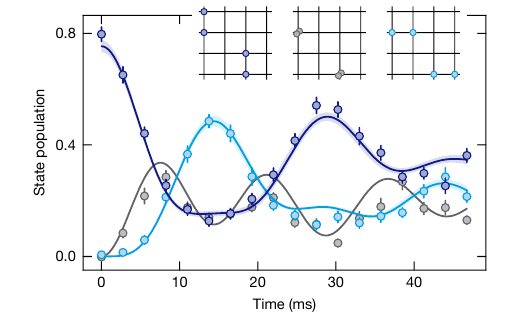}
    \caption{
        \textbf{Dynamics of isolated dimers.}
         Preparing an initial state of dimers with additional free sites, we can study the dynamics of isolated dimer pairs and clearly resolve the oscillatory behavior between the initially prepared vertical (dark blue) and the flipped, horizontal (light blue) dimers. The blue solid lines are fits to the data according to a three state model. Additionally, we can also resolve the occupation of the (grey) doublon state. The grey solid line is the expected doublon occupation for the ideal fit parameters and serves as a guide to the eye. All error bars denote the s.e.m.
    }
    \label{fig:si:iso-dim}
\end{figure}

To further understand and investigate the decay of the dimer pattern on a microscopic level, we prepare isolated dimers and track their evolution after a sudden quench.
We isolate the dimers by adding empty columns between the atom pairs, as illustrated in the inset of Fig.~\ref{fig:si:iso-dim}.
For this configuration, the dimers are, including only first-order processes, completely decoupled from one another, allowing us to study the formation of horizontally-oriented dimers described in Fig.~\ref{fig:3}a in the main text.
The change in orientation can be understood intuitively.
Starting from a dimer, the upper atom can tunnel onto the neighbouring site by forming a doublon, as illustrated in the middle inset of Fig.~\ref{fig:si:iso-dim}.
From there, the atoms can either rearrange into the original dimer or to the flipped dimer, which is energetically degenerate to the original dimer configuration. \hfill \break
Fig.~\ref{fig:si:iso-dim} shows the time evolution of the isolated dimers.
Here, we plot the populations of the three possible states: the vertical dimer, the doublon, and the horizontal dimer.
While the dimer states can be detected unambiguously, we assign the doublon if all three sites are empty.
To correct, on average, for cases where no atoms were initially prepared, we subtract the value obtained analogously from a reference measurement tracking the initial state preparation.
We observe a clear oscillation between the two cases of vertically- and horizontally-oriented dimers, which quickly dephases due to residual potential disorder.
We compare the measured data to a numerical simulation of a three-state model given by

\begin{equation} 
	\hat{H} = \begin{pmatrix} \delta_i & \sqrt{2} J & 0 \\ \sqrt{2} J & U - \Delta + \delta_j & \sqrt{2} J \\ 0 & \sqrt{2} J & \delta_k \end{pmatrix}
\end{equation}
where $\delta_i, \delta_j, \delta_k$ describe the disorder strength between adjacent sites. For the calculation, we sample $\delta_i, \delta_j, \delta_k$ from a normal distribution around zero and average over $N = 100$ such realizations. The additional factor of $\sqrt{2}$ for the hopping has to be taken into account due to the bosonic enhancement characteristic for indistinguishable bosons.
We then fit the calculations to the measured occupation of the vertical, horizontal dimers to generate the solid lines in Fig.~\ref{fig:3}b.
Here, we allow for the disorder strength, the difference $U -\Delta$, an overall amplitude (which respects normalisation) as well as the timescale as free fit parameters. The initial time offset is kept fixed at zero.
Note that the doublon occupation is not included in the fits, instead the solid line in Fig.~\ref{fig:si:iso-dim} is given by the model expectation using the fit values obtained from fitting the two other curves. We observe good agreement between the doublon occupation as obtained from our measured data and the numerical model using the fit parameters for the two other curves, validating our method of extracting the doublon occupation. \hfill \break
From the fit, we extract the standard deviation of the disorder distribution $\sigma = 1.2(1) \times J$, a deviation from resonance of $U - \Delta = 0.0(3) \times J$ and a timescale of $\tau = \SI{10.0(3)}{\milli\second}$. The latter can serve as a secondary way to calibrate the Hubbard parameters of our system (see above).

\subsection{Defects revisited}

\begin{figure}
    \centering
    \includegraphics{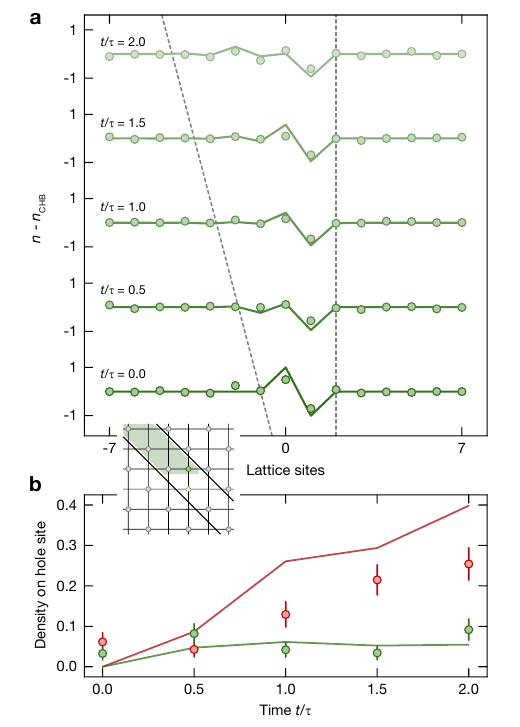}
    \caption{
        \textbf{Additional defect data.}
        \textbf{a} Positive defect on equipotential surface. Spreading of the defect on the zigzag-shaped equipotential line bounded by the black lines shown in the inset. The defect atom can move, but only in one direction, whereas the hole remains immobile. Dashed grey lines are guides to the eye.
        Error bars are on the order of the markers sizes.
        \textbf{b} Density on site initially occupied by the defect hole.
        For the negative defect (red), the hole can move in first order, whereas the hole of the positive defect (green) is stuck.
        Solid lines are theory calculations.
        Error bars denote the s.e.m.
    }
    \label{fig:si:extra-defect-data}
\end{figure}

Here, we describe the data presented in Fig.~\ref{fig:4} and Fig.~\ref{fig:5} in the main text in more detail. We also present an alternative way of evaluating the data for the positive defect and directly compare the spreading of the defect holes for both the negative and positive defect. \hfill \break
The spreading of the defects can be observed directly in the average occupations, see Fig.~\ref{fig:4}b, Fig.~\ref{fig:5}b (leftmost column), through a reduced contrast of the (background) chequerboard on sites accessible to the defect atoms.
This is due to the following processes: first, the defect atoms can move to initially empty sites of the chequerboard, thereby increasing the average density on these sites. 
The defect atoms can also move onto initially occupied sites of the chequerboard, where we then observe a reduced average density due to parity projection.
Finally, nearby atoms from the background chequerboard can become mobile due to the presence of the defect and move onto the site occupied by the defect atom, thus reducing the average density on their original sites as well as on the site of the defect atom due to parity projection.
For the negative defect in particular, the motion of the hole can be observed by an increase of the average occupation on its initial site, see also Fig.~\ref{fig:si:extra-defect-data}b in the following, and a simultaneous decrease of the average occupation on the neighbouring sites on its equipotential line. In contrast, the hole site for the positive defect remains unoccupied.
These effects are highlighted by subtracting the occupation of a chequerboard state without deterministically created defects. 
Initially empty sites accessible to the defect atoms will feature a positive, reference-subtracted value, whereas initially occupied sites accessible to the defect atoms will show negative values. The latter, as explained above, is either due to parity projection, atoms becoming mobile due to the defect, or for the case of the negative defect, also the spreading of the defect hole.
When comparing to theory, we observe good agreement, especially for the negative defect, see Fig.~\ref{fig:4}b and Fig.~\ref{fig:4}c. For the simulations, we did not include any experimental imperfections like disorder and initial state preparation fidelities.
We further quantify the directional spreading of the defects by summing along the diagonals of the reference-subtracted occupations.
When summing parallel to the equipotential lines, the occupation is only different from zero on the diagonals on which the defect atom and hole were initially placed.
The growth by one additional diagonal for times $t / \tau > 0$ can be explained by the above mentioned processes, i.e., the defect's influence on the neighboring atoms.
Note that for these datasets, shown in Fig.~\ref{fig:4}c and Fig.~\ref{fig:5}c, we have rescaled all theoretical calculations to the average value of the experimental data at $t / \tau = 0$ on the two respective, initially non-zero diagonals. \hfill \break
As an additional characterisation of the positive defect, we also studied the spreading on the zigzag-shaped equipotential line (see the inset of Fig.~\ref{fig:si:extra-defect-data}a) instead of summing the reference-subtracted signal along the ROI diagonals.
The result of this analysis is shown in Fig.~\ref{fig:si:extra-defect-data}a.
Here, we again observe that the spread occurs only along one direction, since the immobile hole prevents the spread in the opposite direction.
The latter is expected, as the hole can only move by second-order processes \cite{Will2023}. This is also evidenced by the density on the site of the hole remaining nearly unchanged. \hfill \break
Looking at this further, by comparing the increase of the densities on the sites initially occupied by the defect holes, we can also clearly observe the difference between positive and negative defect, see Fig.~\ref{fig:si:extra-defect-data}b.
For the positive defect, the density increases only slightly, whereas for the negative defect we observe an immediate, fast increase, as here the hole is mobile in first order.
Specifically, the hole of the negative defect can move in processes where neighbouring particles located on the equipotential line above the defect atom hop onto the defect atom and then to the site of the hole, see Fig.~\ref{fig:4}a (bottom right panel).
The hole's motion is restricted to its initial equipotential line.
As for all other measurements on the spreading of defects on top of the chequerboard background, we attribute deviations from the theoretical expectations (no rescaling in Fig.~\ref{fig:si:extra-defect-data}) to disorder in the system and imperfect initial state preparation, i.e., the presence of additional, non-deterministic defects.

\section{Numerical methods}

\subsection{Defects}

\begin{figure}[t!]
    \centering
    \includegraphics[width=1\linewidth]{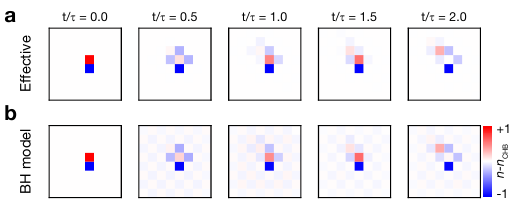}
    \caption{\textbf{Dynamics of the positive defect}. All figures show the parity projected number of particles over time subtracted by a perfect (at $t / \tau = 0$) chequerboard state.
    \textbf{a} Time evolution with the effective Hamiltonian derived in \cite{Will2023} using ED. The background chequerboard pattern is completely frozen, and only the defect is mobile along the diagonal.
    \textbf{b} MPO time-evolution with the full Bose-Hubbard model shows great agreement with the effective time-evolution. In contrast to the effective model, the background chequerboard is not completely frozen. Parameters used: $\mathrm{d}t = 0.0005,\chi = 256, N_{\mathrm{max}} = 2$}    
    \label{fig:SI_num_Def}
\end{figure}

The underlying physics in the Bose-Hubbard model is described by an effective Hamiltonian derived in \cite{Will2023}, which features HSF. 
In Fig.~\ref{fig:SI_num_Def} we compare time evolution of the positive defect under this effective Hamiltonian with time evolution of the original Bose-Hubbard model.
We show the parity projected onsite occupation and have subtracted a perfect chequerboard state (at $t / \tau = 0$, i.e., without time evolution) to better highlight the differences. Note that in Fig.~\ref{fig:4}, Fig.~\ref{fig:5}, and Fig.~\ref{fig:si:extra-defect-data} we instead subtract the theory calculations with a time-evolved version of the chequerboard for better comparison with the experimental data.
In contrast to the effective model, the background chequerboard state is not completely frozen under time evolution with the Bose-Hubbard model. 
Nevertheless, this additional dynamics of the background does not strongly influence the dynamics of the mobile defect as compared to the effective model. 
Therefore, we conclude that the underlying physics of the Bose-Hubbard model in the chosen limits are well captured by the effective model featuring HSF.

\subsection{Convergence}

\begin{figure*}[t!]
    \centering
    \includegraphics[width=1\linewidth]{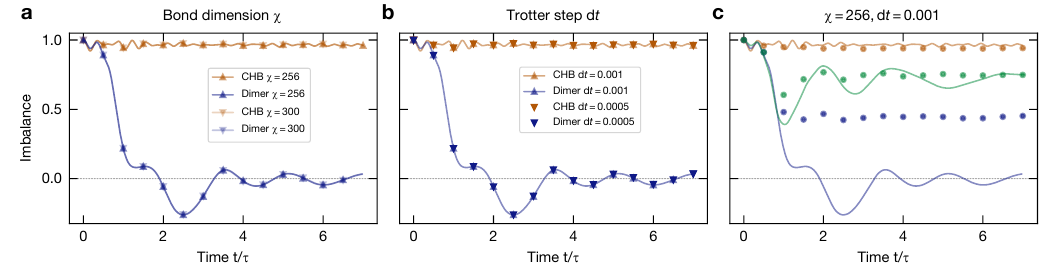}
    \caption{\textbf{Convergence analysis and experimental versus ``clean" time evolution.}
    \textbf{a} Comparison of the time evolution for variable bond dimension $\chi$.
    We find perfect agreement between $\chi = 256$ and $\chi = 300$, demonstrating the convergence of our code, assuming the experimental parameter regime $J = 1, U = \Delta = 18$.
    \textbf{b} Similar analysis as in (a) , yet with variable Trotter step size.
    We again observe excellent agreement.
    \textbf{c} Comparison of a clean and an imperfect system.
    Including experimental imperfections (circles) leads to a less pronounced decrease of the imbalance in comparison to the clean case (solid lines). Even with modified parameters, the effect of state-dependent dynamics is still clearly visible. Parameters used: $J_{\mathrm{x}} = 1, J_{\mathrm{y}} = 1.1 \times J_{\mathrm{x}}, U = \Delta_{\mathrm{x}} = \Delta_{\mathrm{y}}= 18 \times J_{\mathrm{x}}, \break
    \delta = \left[ -2 \times J_{\mathrm{x}}, 2 \times J_{\mathrm{x}} \right], N_{\mathrm{av, Dimer}}\in [29,100], N_{\mathrm{av, squares}}\in [17,100],N_{\mathrm{av, CHB}}\in [10,100], \text{fidelity empty} = 0.97, \text{fidelity empty} = 0.97,  \text{fidelity empty} = 0.97, \text{fidelity filled} = 0.87$. For all simulations, we have assumed a maximal filling of $N_{\mathrm{max}}=2$ per site. The on-site potential $\delta$ is sampled from a uniform distribution of the given width.} 
    \label{fig:SI_I_convergence}
\end{figure*}

Numerical data has been obtained using Tensor Network methods (TN) and Exact Diagonalization (ED). All data has been calculated by Matrix-Product-Operator (MPO) time evolution using the TeNPy package \cite{Hauschild2018}, except time evolution with the effective model in Fig.~\ref{fig:SI_num_Def}, which has been performed with ED. In Fig.~\ref{fig:SI_I_convergence}a,b we show convergence in the bond dimension and in the Trotter step for experimental parameters. In Fig.~\ref{fig:SI_I_convergence}a, the time evolution of the imbalance for the chequerboard and dimer states shows perfect overlap for bond dimensions $\chi = 256$ and $\chi = 300$. For both curves we have used a Trotter step size of $\mathrm{d}t = 0.001$.
In Fig.~\ref{fig:SI_I_convergence}b the imbalance is compared for Trotter step sizes of $\mathrm{d}t = 0.001$ and $\mathrm{d}t = 0.0005$ for $\chi = 256$.

\subsection{Imbalance}

In Fig.~\ref{fig:SI_I_convergence} we compare the imbalance of the clean case to time evolution under imperfect conditions, similar to those of the experiment. 
For the latter, we have included deviations of all relevant quantities away from optimum, fidelities for state preparation, and an additional random onsite potential; see the caption of Fig.~\ref{fig:SI_I_convergence} for details.
Each time step is averaged over $N_{\mathrm{av, Dimer}}\in [29,100], N_{\mathrm{av, squares}}\in [17,100],N_{\mathrm{av, CHB}}\in [10,100]$ different preparations.
We find that the effect of state-dependent dynamics is still clearly visible also for experimental conditions. For the dimer state, the impact of experimental conditions is the most dramatic, which we attribute to the highest sensitivity to imperfect state preparation. In the case of the dimer state, all atoms have only one nearest neighbour. Removing this neighbour directly leads to a decrease in mobility and can induce frozen particles. In contrast, the square state does not suffer from this effect on the same level. Each atom has three nearest neighbours, and therefore one missing neighbour does not lead to frozen sites.

\end{document}